\documentclass[twocolumn,secnumarabic,amssymb,superscriptaddress, nobibnotes, nofootinbib,aps, prd]{revtex4-1}

\usepackage{mathrsfs}
\usepackage{physics}
\usepackage{tikz}
\usepackage[caption=false]{subfig}
\usepackage[colorlinks=true,linkcolor=blue,citecolor=blue,urlcolor=blue]{hyperref}
\usepackage{cleveref}
\usepackage{amsmath,amssymb,amsfonts,amsthm}
\usepackage{graphicx}
\usepackage{ragged2e}
\usepackage{float}
\usepackage{bm}

\def\rar{\rightarrow}

\def\p{\partial}
\def\ra{\rangle}
\def\la{\langle}

\def\no{\nonumber}
\def\bea{\begin{eqnarray}}
\def\eea{\end{eqnarray}}
\def\be{\begin{equation}}
\def\ee{\end{equation}}

\def\p{\partial}
\usepackage{filecontents}

\begin{document}

\title{Quantum Fisher information as a probe for Unruh thermality}%
\author{Jun Feng}%
\email{Corresponding author:\\
j.feng@xjtu.edu.cn}
\affiliation{School of Physics, Xi'an Jiaotong University, Xi'an, Shaanxi 710049, China}
\affiliation{Institute of Theoretical Physics, Xi'an Jiaotong University, Xi'an, Shaanxi 710049, China}

\author{Jing-Jun Zhang}
\affiliation{School of Physics, Xi'an Jiaotong University, Xi'an, Shaanxi 710049, China}


%
\date{\today}%

\begin{abstract}

A long-standing debate on Unruh effect is about its obscure thermal nature. In this Letter, we use quantum Fisher information (QFI) as an effective probe to explore the thermal nature of Unruh effect from both local and global perspectives. By resolving the full dynamics of UDW detector, we find that the QFI is a time-evolving function of detector's energy gap, Unruh temperature $T_U$ and particularities of background field, e.g., mass and spacetime dimensionality. We show that the asymptotic QFI whence detector arrives its equilibrium is solely determined by $T_U$, demonstrating the global side of Unruh thermality alluded by the KMS condition. We also show that the local side of Unruh effect, i.e., the different ways for the detector to approach the same thermal equilibrium, is encoded in the corresponding time-evolution of the QFI. In particular, we find that with massless scalar background the QFI has unique monotonicity in $n=3$ dimensional spacetime, and becomes non-monotonous for $n\neq3$ models where a local peak  value exists at early time and for finite acceleration, indicating an enhanced precision of estimation on Unruh temperature at a relative low acceleration can be achieved. Once the field acquiring mass, the related QFI becomes significantly robust against the Unruh decoherence in the sense that its local peak sustains for a very long time. While coupling to a more massive background, the persistence can even be strengthened and the QFI possesses a larger maximal value. Such robustness of QFI can surely facilitate any practical quantum estimation task.


\end{abstract}

\maketitle

\section{Introduction}
\label{1}

The celebrated Unruh effect predicts that an observer with uniform acceleration $a$ may perceive the ordinary Minkowski vacuum as a thermal state of temperature $T_U=a/2\pi$ \cite{UNRUH1}. It manifests the observer-dependence of the particle content of quantum fields, which happens naturally in various scenarios, such as Hawking radiation of black hole \cite{UNRUH2} or particle creation in inflationary universe \cite{UNRUH3}. Thus, aside from its own interest, Unruh effect can also shed light on yet unknown issue of quantum gravity that are much less tractable (see \cite{UNRUH4} for a recent review).

The Unruh effect can be explicitly revealed by the response spectrum of a moving Unruh-DeWitt (UDW) detector that couples to the field modes of interest \cite{UNRUH5}. In $n=4$ dimensional flat spacetime, once being in equilibrium with a massless scalar background, the power spectrum of the accelerating detector exhibits a Planckian form, agreeing precisely with the conventional thermal spectrum in a heat bath, therefore strongly suggests the interpretation of $T_U$ as a temperature. In an axiomatic way, it has been proven \cite{UNRUH6,UNRUH7} that the Unruh effect is universal in arbitrary dimensions independently of whether or not particles are interacting. 

Nevertheless, the thermal nature of Unruh effect is obscure and remains puzzles. Firstly, an oft-heard explanation on the thermal character of Unruh radiation relies on so-called thermalization theorem \cite{THERMAL1,THERMAL2}, which indicates that an accelerating detector is causally disconnected from the degrees of freedom screened by the \emph{global} Rindler horizon, causing an information loss presented by a Kubo-Martin-Schwinger (KMS) state, described by a canonical ensemble. However, this does not naively lead to an observed detector's spectrum in Planck form. For example, within a massive scalar background, the response rate of detector can be far from the form with Planckian factor. Indeed, the thermalization theorem only guarantees the KMS condition \cite{TAKAGI1}, or equivalently the principle of detailed balance, and nothing more. In this light, instead of detector's response spectrum, various measures of quantum correlation across the horizon, such as entanglement entropy \cite{ee1,ee2}, quantum discord \cite{Discord1,Discord2}, uncertainty relation \cite{F2,F3} and nonlocality \cite{BF3,F5}, have been employed to encapsulate the \emph{global} trait of Unruh thermality.

In generic $n$-dimensional spacetime, the detector's local response may also exhibit an anomalous ''statistics inversion'', an apparent interchange between Bose-Einstein and Fermi-Dirac distributions of the spectrum \cite{TAKAGI2,TAKAGI3}. In particular, for an UDW detector coupled to a massless scalar field, Takagi shown \cite{TAKAGI2} that the excited power spectrum obeys Bose-Einstein distribution with even $n$, but turns to be Fermi-Dirac for odd $n$, even without any fermionic degrees of freedom existing. Later pointed in \cite{TAKAGI4}, this was ascribed to the absence of Huygens principle in odd dimensional spacetime, a result of the analytic structure of Wightman function, leading to anticommutativity for timelike separated scalar correlator. Recent works further argue \cite{TAKAGI5,TAKAGI6} that the spurious statistics inversion should be viewed as a \emph{local} feature of the detector, encoding the nontrivial density of states on the Rindler wedge, or the way of the detector approaching its thermal equilibrium \cite{TAKAGI7}. Although such particularities of the response spectrum would not spoil the thermal character of the Unruh effect, which is solely determined by the KMS condition, it is interesting to ask \cite{TAKAGI8} if any physical meaningfulness can be extracted from those local response.

In this Letter, we use quantum Fisher information (QFI), a key concept in both quantum information theory and metrology, to explore the thermal nature of Unruh effect from both local and global perspectives. In particular, we estimate the QFI for an UDW detector, linearly coupled to scalar field background in arbitrary dimensional flat spacetime. We treat the UDW detector as an open quantum system \cite{OP1}, whose dynamics can then be derived from the time evolution of combined system (detector+external field) by integrating over the field degrees. In terms of its response function, the density matrix of detector can be explicitly resolved, which asymptotically reaches a thermal mixed state \cite{OP2,OP3}. This approach enables us to investigate Unruh effect from a refined perspective, rather than the asymptotic excitation rate of the detector only. 
 
Our interests in using QFI to diagnose Unruh effect are three-fold. Firstly, as an operational measure on the distinguishability of quantum states, Fisher information can discriminate parameterized states $\rho({X})$ and $\rho(X+\delta X)$ with infinitesimal change of parameter $X$. We will argue that, given the time-dependent state of UDW detector, the Fisher information $\mathcal{F}_X(\rho)$ may encode the particularities of detector's response function, e.g., those inherited from the specific coupling ways \cite{FI1,FI5} or the statistics inversion in different spacetime dimensions, therefore reveals the local trait of the Unruh effect. Secondly, as a consequence of the Cram\'er-Rao inequality, the QFI quantifies the highest-precision of parameter measurement achieved in a metrological scheme. For $\theta$ as the parameter governing relativistic phenomena, intensive works have been made on utilizing QFI to probe extremely sensitive quantum gravity effects \cite{FI2,FI3,FI3+} or spacetime structure \cite{FI4,FI5,FI6}. In our scheme, by choosing $X=\beta\equiv1/T_U$, the QFI bounds the metrological accuracy of any quantum estimation on Unruh temperature. By efficiently designing an UDW detector, we may significantly enhance the QFI and preserve it against environment noise. Finally, since an entangled state provides a higher precision than an ensemble of uncorrelated particles, the inverse of QFI can serve as a \emph{genuine} entanglement probe, based on the metrological performance of the quantum states \cite{FI7,FI8}. This suggests that once obtaining the QFI of UDW detector, we can find the imprint of the non-local nature of Unruh effect encoding the quantum correlations across the Rindler horizon \footnote{It is interesting to note \cite{FI9,FI10} that such connection between Fisher information and entanglement has recently been identified in holographic picture, where the holographic dual of QFI near vacuum in CFT gives bulk entanglement and canonical energy.}, as well as its surviving in a noisy environment. Recall that the entangled side of Unruh effect is directly related to the universal thermalization, and has nothing to do with having a strictly Planckian response function. This may seem in conflict to the fact that QFI behaves differently for the same accelerating detector with different local responses. Nevertheless, we will show that the universal thermal nature of Unruh effect can be represented by the asymptotically convergent of QFI, which is ascribed to the KMS condition \cite{FI5}, for times much longer than the scale of thermalized process.

The Letter is organized as follows. In Section \ref{2}, we solve the master equation of the UDW detector as an open quantum system in $n$-dimensional flat spacetime, and give explicitly its time-dependent state. In Section \ref{3}, we evaluate the QFI of UDW detector, whose time evolution encodes the local response in different scenarios and asymptotically convergent represents the universal thermal nature of Unruh effect. In Section \ref{4}, the summary and discussion are given. Throughout this Letter, we use units with $G=c=\hbar=k_{B}=1$.

\section{Dynamical evolution of UDW detector in $n$-dimensional spacetime}
\label{2}

\subsection{Master equation and KMS condition}
To proceed, we should first explore the full dynamics of an accelerating UDW detector in $n$-dimensional Minkowski spacetime. Treating the detector as an open quantum system coupled to a bath of fluctuating quantum scalar field, its density matrix is governed by a master equation in Lindblad form and evolves non-unitarily, due to the environment decoherence and dissipation on the system.

The total Hamiltonian of the combined system of detector and quantum field is
\be
H=H_{\text{atom}}+H_\Phi+\mu H_I             \label{eq1}
\ee
where UDW detector is modeled by a two-level atom $H_{\text{atom}}=\frac{1}{2}\omega \sigma_3$, $H_\Phi$ is the Hamiltonian of free scalar field $\Phi(x)$, which satisfies the standard Klein-Gordon equation in $n$-dimensional Minkowski spacetime. The atom-field interaction is given by $H_I=(\sigma_++\sigma_-)\Phi(x)$. Here, we denote the atomic raising and lowering operators as $\sigma_\pm$, and $\omega$ is the energy level spacing of the atom. 

Assuming the detector and environment are weakly coupled, such that the initial state of combined system can be approximated as $\rho_{tot}(0)=\rho(0)\otimes |0\ra\la0|$, where $\rho(0)$ is the initial state of the atom and $|0\ra$ is the field vacuum respecting spacetime symmetry. Obviously, the dynamics of $\rho_{tot}$ should be governed by a unitary evolution via von Neumann equation $\dot{\rho}_{tot}(\tau)=-i[H,\rho_{tot}(\tau)]$, where $\tau$ is the proper time of the detector. Whenever the typical time scale of the environment is much smaller than that of the detector, we can further assume it is under a Markovian evolution. The reduced dynamics of the detector is then obtained by integrating over the background field degrees from the $\rho_{tot}(\tau)$, governed by a quantum dynamical semigroup of completely positive map\footnote{The Markov approximation enables the upper limit of the integral can be lifted to infinite proper time, which is permissible as the integrand disappear for the large time over which the field correlation function decay sufficiently fast. This leads the Kossakowski coefficients (\ref{eq6}) to be integrated over an infinite range of detector's proper time \cite{OP1}.}, and is represented by a Kossakowski-Lindblad master equation \cite{OP4,OP5}
\be
\frac{d\rho}{d\tau}=-i\left[H_{\mbox{\tiny eff}},\rho(\tau)\right]+\mathcal{L}\left[\rho(\tau)\right]             \label{eq2}
\ee
where
\be
\mathcal{L}\left[\rho\right]=\sum^3_{i,j=1}C_{ij}\left[\sigma_j\rho \sigma_i-\frac{1}{2}\left\{\sigma_i\sigma_j,\rho\right\}\right]        \label{eq3}
\ee
represents a dissipative evolution attributed to the interaction between the detector and external fields. The Kossakowski matrix $C_{ij}$ can be explicitly resolved. After introducing the Wightman function of scalar field $G^{+}\left(x, x^{\prime}\right)=\left\langle 0\left|\Phi(x) \Phi\left({x}^{\prime}\right)\right| 0\right\rangle$, its Fourier transform
\be
\mathcal{G}(\lambda)=\int_{-\infty}^{\infty}d\tau~e^{i\lambda\tau}G^+(\tau)= \int_{-\infty}^{\infty}d\tau~e^{i\lambda\tau}\left\la\Phi(x(\tau))\Phi(x(0))\right\ra                 \label{eq4}
\ee
determines the coefficients $C_{ij}$ through the decomposition
\be
C_{ij}=\frac{\gamma_+}{2}\delta_{ij}-i\frac{\gamma_-}{2}\epsilon_{ijk} n_k+\gamma_0\delta_{3,i}\delta_{3,j}          \label{eq5}
\ee
where
\be
\gamma_\pm=\mathcal{G}(\omega)\pm \mathcal{G}(-\omega),~~~\gamma_0=\mathcal{G}(0)-\gamma_+/2         \label{eq6}
\ee
up to an overall factor $\mu^2$ from detector-field coupling, which usually can be set to 1 for the sake of simplicity. Moreover, the interaction with external scalar field would also induce a Lamb shift contribution for the detector effective Hamiltonian $H_{\mbox{\tiny eff}}=\frac{1}{2}\tilde{\omega}\sigma_3$, in terms of a renormalized frequency $\tilde{\omega}=\omega+i[\mathcal{K}(-\omega)-\mathcal{K}(\omega)]$, where $\mathcal{K}(\lambda)=\frac{1}{i\pi}\mbox{P}\int_{-\infty}^{\infty}d\omega\frac{\mathcal{G}(\omega)}{\omega-\lambda}$ is Hilbert transform of Wightman functions. 

For later purposes, we can rewrite (\ref{eq5}) in a more instructive form. Following a trajectory of the accelerating detector, one can find that the field Wightman function fulfills the KMS condition, i.e., $G^{+}(\tau)=G^{+}(\tau+i \beta)$. Translating it into frequency space, one has 
\be
\mathcal{G}(\lambda)=e^{\beta\omega}\mathcal{G}(-\lambda)        \label{eq7}
\ee
Further using the translation invariance $\langle 0|\Phi(x(0)) \Phi(x(\tau))| 0\rangle=\langle 0|\Phi(x(-\tau)) \Phi(x(0))| 0\rangle$, after some algebras, we find that (\ref{eq6}) can be represented as
\bea
\gamma_+&=&\int_{-\infty}^{\infty}d\tau~e^{i\lambda\tau}\la0|\left\{\Phi(\tau),\Phi(0)\right\}|0\ra=\left(1+ e^{-\beta\omega}\right) \mathcal{G}(\omega),\no\\
\gamma_-&=&\int_{-\infty}^{\infty}d\tau~e^{i\lambda\tau}\la0|\left[\Phi(\tau),\Phi(0)\right]|0\ra=\left(1- e^{-\beta\omega}\right) \mathcal{G}(\omega),\no\\
\label{eq8}
\eea
holding true for generic interacting fields with or without mass. 

It also useful to introduce the ratio
\be
\gamma\equiv\gamma_-/\gamma_+=\frac{1- e^{-\beta\omega}}{1+ e^{-\beta\omega}}=\tanh(\beta\omega / 2)        \label{eq9}
\ee
which depends only on the Unruh temperature $T_U=1/\beta$ due to the frequency KMS condition (\ref{eq7}), but has nothing to do with local correlator of the background.

For the UDW detector modeled by a two-level atom, its density matrix can be expressed in a Bloch form
\be
\rho(\tau)=\frac{1}{2}\Big(1+\sum_{i=1}^3\rho_i(\tau)\sigma_i\Big)        \label{eq10}
\ee 
Substituting it into the master equation (\ref{eq3}), the Bloch vector $\bm{\rho}=(\rho_1,\rho_2,\rho_3)^T$ satisfies \cite{OP6}
\be
\frac{\partial \bm{\rho}}{\partial \tau}+2 \bm{K}\cdot\bm{\rho}+\bm{\eta}=0          \label{eqplus-1}
\ee
where $\bm{\eta}=(0,0,-2\gamma_-)^T$ and $\bm{K}$ is
\be
\bm{K}=\left(\begin{array}{ccc}
 \gamma_++\gamma_0 & {\tilde{\omega}} / 2 & 0 \\
-{\tilde{\omega}} / 2 &  \gamma_++\gamma_0 & 0 \\
0 & 0 &  \gamma_+
\end{array}\right)            \label{eqplus-2}
\ee
Assuming a general initial state $|\psi\ra=\sin\frac{\theta}{2}|0\ra+\cos\frac{\theta}{2}|1\ra$, the master equation (\ref{eq2}) can be analytically resolved. The time-dependent components of the state are given as
\bea
\rho_1(\tau)&=&e^{-\frac{1}{4}\gamma_+\tau}\sin\theta\cos\tilde{\omega}\tau\no\\
\rho_2(\tau)&=&e^{-\frac{1}{4}\gamma_+\tau}\sin\theta\sin\tilde{\omega}\tau\no\\
\rho_3(\tau)&=&e^{-\frac{1}{2}\gamma_+\tau}\left(\cos\theta+\gamma\right)-\gamma            \label{eq11}
\eea

It is easy to see that for large time $\tau\gg1$, the detector's state reaches at a thermal equilibrium with Unruh temperature like
\be
\rho_{\text{equil}}(\beta)=\frac{1}{2}\left(\begin{array}{cc}1-\gamma & 0 \\0 & 1+\gamma\end{array}\right)=\frac{e^{-\beta H_{\text{atom}}}}{\operatorname{Tr}\left[e^{-\beta H_{\text{atom}}}\right]}           \label{eq12}
\ee
An important insight \cite{TAKAGI7} is that even for the same end of thermalization, the detector can follow different paths within its space of states (\ref{eq11}) to reach thermal equilibrium, parameterized by its local features such as the initial state preparation and the spacetime dimensionality.

With the reduced density matrix (\ref{eq11}), the full information of the nonunitary evolution of the atom can be extracted. A  best known example is the transition rate of the detector between its initial and final states, i.e., $\mathcal{P}_{i \rightarrow f}(\tau)=\operatorname{Tr}\left[\rho_{f} \rho(\tau)\right]$. When initially $\rho(0)\equiv\rho_{i}$ and $\rho_{f}$ are the ground and excited states of the detector, i.e., the eigenstates of the Hamiltonian $H_{\text{atom}}=\frac{1}{2}\omega \sigma_3$ in (\ref{eq1}), then the probability for a spontaneous transition of the detector is \cite{OP2}
\be
\mathcal{P}_{\uparrow}=\frac{\gamma_+-\gamma_-}{2 \gamma_+}\left(1-e^{-2 \gamma_+ \tau}\right) .
\ee
which gives the transition probability per unit time at $\tau=0$ as
\be
\Gamma_{\uparrow}=\left.\frac{\partial \mathcal{P}_{\uparrow}}{\partial \tau} \right|_{\tau=0}=\mathcal{G}\left(-\omega\right)
\ee
This spontaneous excitation rate is what usually referred as the Unruh effect. We note that $\Gamma_{\uparrow}$ does not contain the renormalized frequency $\tilde{\omega}$ since it indeed manifests the detection rate of the Rindler quanta from thermalized Minkowski vacuum, while the atom's Lamb shift is responsible for the AC Stark shift associated with these quanta \cite{OP7}.   

\subsection{State dynamics with various responses}

To derive the complete dynamics of detector's state (\ref{eq11}), we need to specify $\mathcal{G}(\omega)$ following the trajectory of atom with uniformly acceleration, like
\be
x^0(\tau) =a^{-1} \sinh a \tau, ~ x^1(\tau)=a^{-1} \cosh a \tau ,~\bm{x}(\tau)=\text{const}           \label{eq13}
\ee
in $n$-dimensional Minkowski spacetime. Expending the free scalar field as
\be
\Phi(x)=\int d^{n-1} k\left\{a_{k} U_{k}(x)+{a}_{k}^{+} U_{k}^{*}(x)\right\}           \label{eq14}
\ee
the positive field mode can be read as
\be
U_{k}=\left[2 \Omega_{k}(2 \pi)^{n-1}\right]^{-1 / 2} \exp i\left(k_{1} x^{1}+\boldsymbol{k} \cdot \boldsymbol{x}-\Omega_{k} x^{0}\right)           \label{eq15}
\ee
where $\Omega_{k}^2=m^2+k_1^2+|\bm{k}|^2$ and the creation/annihilation operators obey canonical communication relation $[a_{k}, a_{k^{\prime}}^{\dag}]=\delta^{n-1}\left(k-k^{\prime}\right)$.

Substituting (\ref{eq14}) and (\ref{eq13}) into (\ref{eq4}), for an accelerating UDW detector, we can formally obtain
\be
\mathcal{G}_n(\omega)=-\frac{\pi}{\omega} \frac{D_{n}(\omega) }{e^{- \beta\omega }-1}     \label{eq16}
\ee
where the profile function $D_{n}(\omega)$ is given by an integral of modified Bessel function \cite{TAKAGI1}
\be
D_n(\omega)=\frac{2}{\pi|\Gamma(i \omega / a)|^{2}} \int \frac{d^{n-2} k}{(2 \pi)^{n-2}}\left|K_{i \omega / a}\left(\sqrt{m^2+|\bm{k}|^2}/ a\right)\right|^{2}          \label{eq17}
\ee

For a free massless scalar field in $n$-dimension, (\ref{eq16}) has an analytical form
\be
\mathcal{G}^{m=0}_n(\omega)=\frac{ \pi^{\frac{n-5}{ 2}}\beta^{3-n}}{2\Gamma\left(\frac{n-1}{ 2}\right)}\left|\Gamma\left(\frac{n}{2}-1+ \frac{\beta\omega}{2\pi}i\right) \right|^{2}\frac{f_{n}(\omega)}{e^{-\beta\omega}-(-1)^{n}}         \label{eq18}
\ee
with
\be
f_{n}(\omega)=\left\{\begin{array}{ll}
-\operatorname{sinh}(\beta\omega / 2) & \text { if } n \text { is even } \\
\operatorname{cosh}(\beta\omega / 2) & \text { if } n \text { is odd }
\end{array}\right.         \label{eq19}
\ee
We observe a Plankian factor with Bose-Einstein distribution for even $n$, but Fermi-Dirac distribution for odd $n$, that clearly manifests Takagi's statistics inversion. This phenomenon also convinces that Unruh effect is essentially distinct from a simple thermal noise. To see it, one only need to evaluate the Wightman function in a thermal state with Unruh temperature $T_U$ \cite{TAKAGI2}, i.e. $\left\langle \Phi(x) \Phi\left({x}^{\prime}\right)\right\rangle_\beta$, from which we evaluate (\ref{eq4}) as
\be
\mathcal{G}_{thermal}\sim-\frac{2^{2-n} \pi^{\frac{1-n}{ 2}}\omega^{n-3}}{\Gamma\left(\frac{n-1}{ 2}\right)}\frac{1}{e^{-\beta\omega }-1}            \label{eq19++}
\ee
Apparently, the statistics perceived by the detector remains in arbitrary spacetime dimensions.

For a free massive scalar field in $n$-dimension, unfortunately, the integral (\ref{eq17}) admits no simple analytic expression. Nevertheless, we can alternatively evaluate this asymptotically \cite{TAKAGI1}, which will be sufficient for our concern on the possible mass effect encoded in QFI. In particular, for large field mass with $m\gg T_U$, using the asymptotic form of the modified Bessel function for large argument, the profile function $D_{n}(\omega)$ can be evaluated, which gives (\ref{eq16}) like
\be
\mathcal{G}^{\text{massive}}_n(\omega)\approx\frac{m^{n / 2-2} e^{-m\beta / \pi}}{2^{n/2} \beta^{n/2-1}} e^{\beta\omega / 2}      \label{eq21}
\ee
Although does not contain Planck factor in an explicit way,  (\ref{eq21}) is still thermal in the sense that it fulfills the frequency KMS condition, i.e., $\mathcal{G}^{\text{massive}}_n(-\omega)=e^{-\beta\omega}\mathcal{G}^{\text{massive}}_n(\omega)$.

Substituting (\ref{eq18}) and (\ref{eq21}) into (\ref{eq8}), one can obtain the Kossakowski coefficients for scalar background with large mass
\be
\gamma_{+,~n}^{\text{massive}}=\frac{m^{n / 2-2} e^{-m\beta / \pi}}{2^{n/2-1} \beta^{n/2-1}} \operatorname{cosh}(\beta\omega / 2)      \label{eq22}
\ee
and for massless scalar field
\be
\gamma_{+,~n}^{m=0}=\frac{ \pi^{\frac{n-5}{ 2}}\beta^{3-n}}{2\Gamma\left(\frac{n-1}{ 2}\right)}\left|\Gamma\left(\frac{n}{2}-1+ \frac{\beta\omega}{2\pi}i\right) \right|^{2} \operatorname{cosh}(\beta\omega / 2)       \label{eq23}
\ee
Using Euler's reflection formula $|\Gamma(1 / 2+i x)|^{2}=\pi / \cosh (\pi x)$ and recurrence relation $\Gamma(z+1)=z \Gamma(z)$, we give the first several cases of (\ref{eq23}) as
\bea
\gamma_{+,~3}^{m=0}=\frac{1}{2}&,&\gamma_{+,~4}^{m=0}=\frac{\omega}{2\pi}\gamma^{-1}\no\\
\gamma_{+,~5}^{m=0}=\frac{\pi}{8\beta^2}+\frac{\omega^2}{8\pi}&,&\gamma_{+,~6}^{m=0}=\left(\frac{\omega}{3\beta^2}+\frac{\omega^3}{12\pi^2}\right)\gamma^{-1}
     \label{eq24}
\eea
for later use. The hyperbolic factor $\gamma^{-1}$ appears only for even $n$, inherited from the statistics inversion.

We note that dimension $n=3$ is quite unique in the sense that its Kossakowski coefficient is a constant, while those for other dimension are modulated by a polynomial on $\beta$. Later, we will see that it is such speciality of $n=3$ making the derived QFI evolves drastically different from the models with other dimension.

\section{QFI for UDW detector and Unruh thermality}
\label{3}

Our aim is to explore the QFI of detector's state (\ref{eq11}), with Unruh temperature $\beta=1/T_U$ as a parameter chosen to be optimally estimated. We anticipate that the different time evolution of QFI can manifest the local feature of Unruh effect, ascribed to various detector's response under the interaction with scalar background. Moreover, within a metrological task on estimating Unruh temperature, QFI bounds the maximal precision of estimator. Therefore, how to effectively enhance the QFI by properly design an UDW detector, e.g., carefully selecting its initial state, level spacing as well as its interacting way to the background, may also be revealed for the particular cases fixed by (\ref{eq22}) and (\ref{eq23}). Finally, as the equilibrium state (\ref{eq12}) depends solely on Unruh temperature, we would naturally expect that asymptotic behavior of QFI should represent the thermal nature of Unruh effect which is guaranteed by the KMS condition alone.

\subsection{Fisher information for UDW detector}
In any quantum metrological tasks, QFI gives a lower bound to the mean-square error in the parameter estimation via the Cram\'er-Rao inequality $\operatorname{Var}(X) \geqslant\left[N \mathcal{F}_{X}\right]^{-1}$, where $N$ is the number of repeated measurements. In terms of the symmetric logarithmic derivative (SLD) operator $L_X$, which satisfies $\p_X\rho=\frac{1}{2}\{\rho,L_X\}$, the QFI is defined as $\mathcal{F}_X=\mbox{Tr}[\rho(X)L^2_X]$. For a diagonalized density matrix like $\rho=\sum_{i=\pm}\lambda_i|\psi_i\ra\la\psi_i|$, the related QFI can be further written as \cite{FI11,FI12}
\be
\mathcal{F}_X=\sum_{i=\pm}\frac{(\p_X \lambda_i)^2}{\lambda_i}+\sum_{i\neq j=\pm}\frac{2(\lambda_i-\lambda_j)^2}{\lambda_i+\lambda_j}|\la\psi_i|\p_X\psi_j\ra|^2            \label{eq25}
\ee
where the summations run over all eigenvalues satisfying $\lambda_i\neq0$ and $\lambda_i+\lambda_j\neq0$.

In our context, by diagonalizing the detector's density matrix (\ref{eq10}), the related QFI on parameter $\beta=1/T_U$ for arbitrary initial state can be straightforwardly calculated as \cite{FI5}
\be
\mathcal{F}_\beta=\frac{\left(\partial_{\beta}\ell\right)^{2}}{1-\ell^{2}}+\frac{\left(\p_\beta(\rho_3/\ell) \right)^{2}}{1-(\rho_3/\ell)^2}+\tau^{2}\left(\rho_{1}^{2}+\rho_{2}^{2}\right)\left(\partial_{\beta} \tilde{\omega}\right)^{2}             \label{eq26}
\ee
where $\ell\equiv\sqrt{\sum_{i}\rho_{i}^{2}}$ is the length of Bloch vector. For given mixed state (\ref{eq11}), the length of Bloch vector should satisfy $\ell<1$, which means each term of (\ref{eq26}) is positive and be periodic in $\theta$. By varying $\theta$ only,  it is not hard to find \cite{FI5} that the QFI can achieve its maximum once the detector be prepared in the initial state with $\theta=(2 k+1) \pi$ $(k \in \mathbb{Z})$. 

The full dynamics of an UDW detector in $n$-dimensional flat spacetime indicates that the QFI is a time-evolving function as $\mathcal{F}_{\beta}\left(\tau,\omega,\theta;n\right)$, which relays on the initial state preparation fixed by $\theta$, the energy level spacing of atom $\omega$, as well as the particularities of the background field, such as its mass and spacetime dimensionality $n$, encoded in detector's local response. To access the highest precision of estimation, we have to maximize the value of QFI over all relevant parameters it depends on. 

For simplicity, we can choose the initial state with $\theta=\pi$, i.e., the detector is initially prepared in ground state $|0\ra$, which gives $\ell=|\rho_3|$. By straightforwardly calculation on (\ref{eq26}), we obtain
\be
\mathcal{F}_{\beta}\left(\tau,\omega;n\right)=\frac{\left[
\left(1-e_n\right)\p_\beta\gamma -\frac{\tau}{2}e_n(1-\gamma)\p_\beta\gamma_{+,n}\right]^{2}}{1-\left[e_n+\gamma(1-e_n)\right]^2}             \label{eq28}
\ee
with abbreviation ${e}_n\equiv \exp\left(-\gamma_{+,n} \tau/2\right)$. This expression would be used later to analyze the QFI in various scenarios.

In later discussion, we will work with dimensionless parameters by rescaling the Unruh temperature and field mass apparently as
\be
\beta \longmapsto \tilde{\beta} \equiv \beta \omega,\quad m\longmapsto \tilde{m} \equiv m/\omega
\ee
Particularly cautious is needed when we rescale the proper time to a dimensionless parameter. Note that the stationary state (\ref{eq12}) is reached asymptotically via the exponential decay of the $\tau$-dependent term in (\ref{eq11}). This means that $\gamma_+^{-1}$ has a characteristic time scale for arbitrary dimension, and its combination with $\tau$ gives a dimensionless parameter. Therefore, we define a dimensionless time parameter as\footnote{This may seems strange since only $\mu_4$ has the dimension $[\text{Time}]^{-1}$. However, the rescaling should actually be $\mu_n=\mu^2\omega^{n-3}$ where the detector-field coupling $\mu^2$ has been absorbed before. For $n\neq 4$, the coupling constant is dimensionful to maintain the combination $\gamma_+\tau$ dimensionless \cite{TAKAGI7}.}
\be
\tau \longmapsto \tilde{\tau} \equiv \mu_{n} \tau, \quad 
\ee
where $\mu_n=\omega^{n-3}$. However, for convenience, we continue to term $\tilde{\beta}$, $\tilde{\tau}$ and $\tilde{m}$, respectively, as $\beta$, $\tau$ and $m$.

\subsection{Asymptotic QFI and KMS condition}
\begin{figure}[hbtp]
\begin{center}
\includegraphics[width=.4\textwidth]{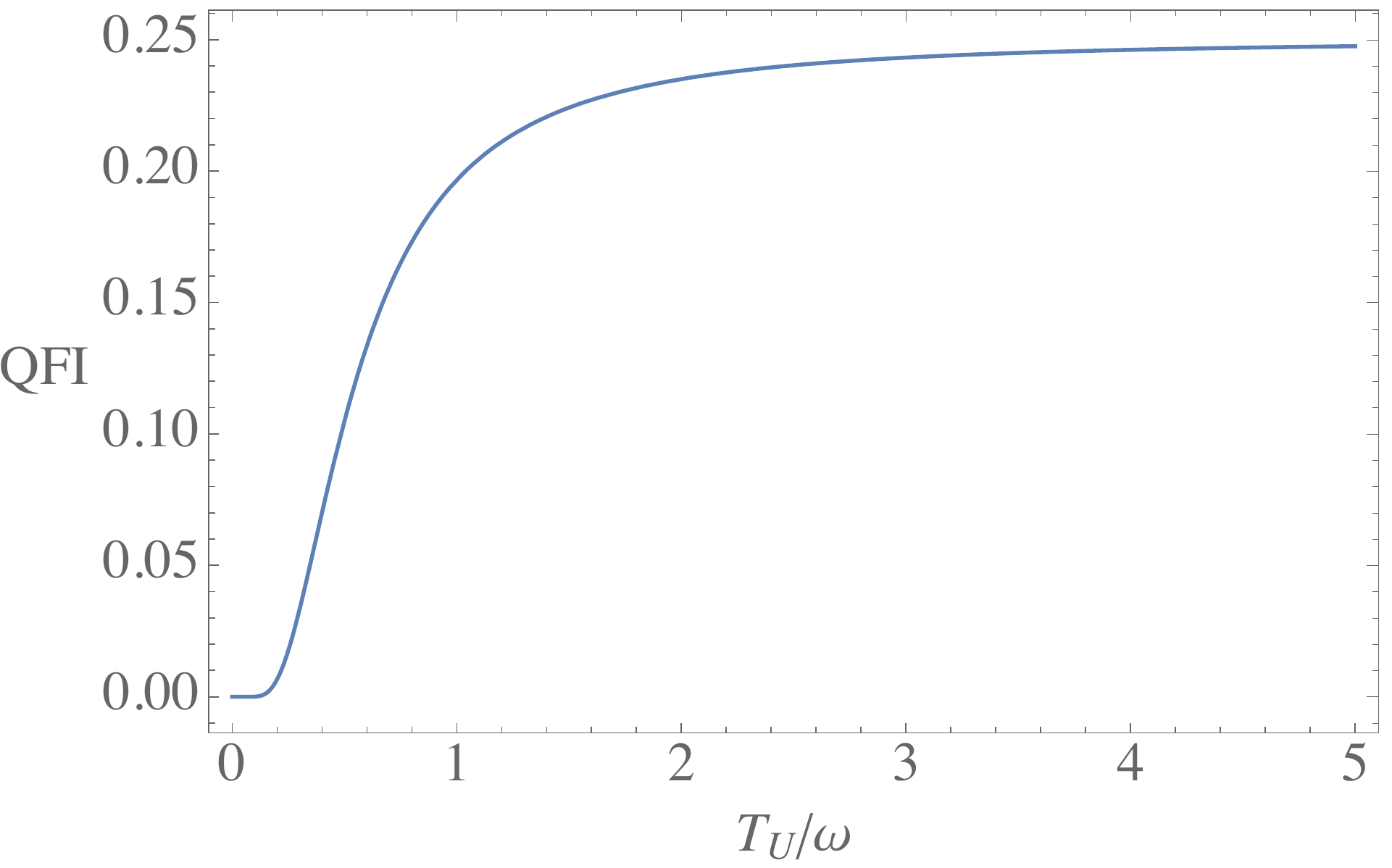}
\caption{The QFI for the asymptotic equilibrium state of an accelerating UDW detector, which is independent on the its local response to the background and represents the global thermal nature of Unruh effect.}
\label{fig1}
\end{center}
\end{figure}

After evolving for enough long time, the QFI (\ref{eq26}) of the reached equilibrium state (\ref{eq12}) has a simple form
\be
\mathcal{F}_{\beta}\Big|_{\tau\rar\infty}=\frac{1-\gamma^2}{4}       \label{eq27}
\ee
as depicted in Fig.\ref{fig1}. One can observe that the asymptotic QFI has nothing to do with the background field, i.e., be irrelevant to the local response of the detector, therefore representing the global thermal nature of Unruh effect.  

As Unruh temperature growing, the QFI value increases which means higher precision of parameter-estimation can be achieved for the detector with larger acceleration. This is reasonable in the sense that it is always easy to figure a distinctive signal, e.g., high $T_U$, out from the average background. In a combined limit of infinite acceleration and long evolving time, we have the asymptotic QFI as $\mathcal{F}_{\text{asym}}\left(\beta\rar0\right)=0.25$.    

\subsection{QFI and spacetime dimensionality}
We now estimate the QFI of an accelerating detector within massless field background. For arbitrary dimensional flat spacetime, the detector's response function exhibits a statistics inversion manifested in (\ref{eq18}). Our aim now is to exploit that how spacetime dimensionality may influence the QFI, through its evolution in various dimension.

\begin{figure}[hbtp]
\begin{center}
\subfloat[$\mathcal{F}_{\beta}(n=3)$]{\includegraphics[width=.235\textwidth]{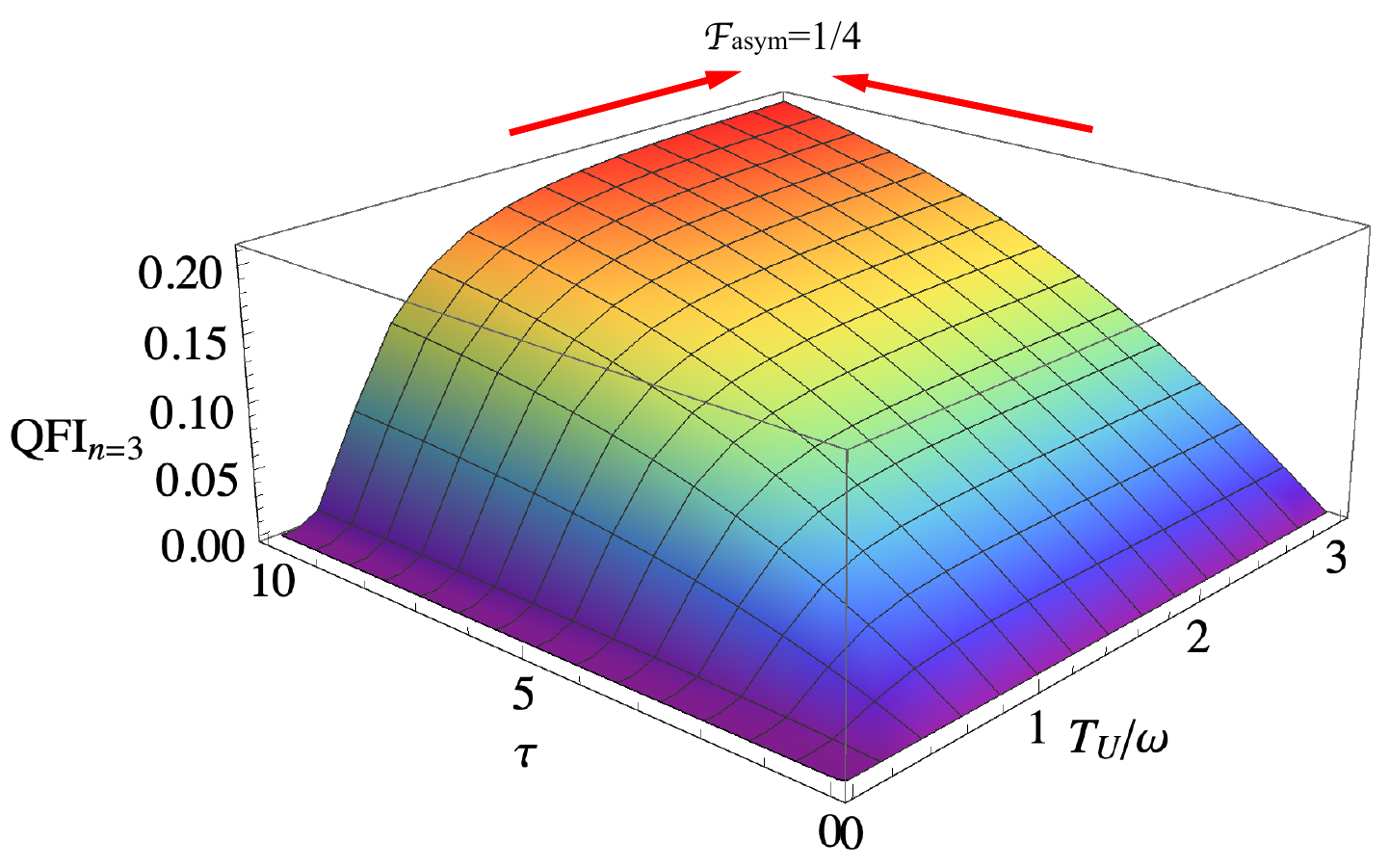}}
\subfloat[$\mathcal{F}_{\beta}(n=4)$]{\includegraphics[width=.245\textwidth]{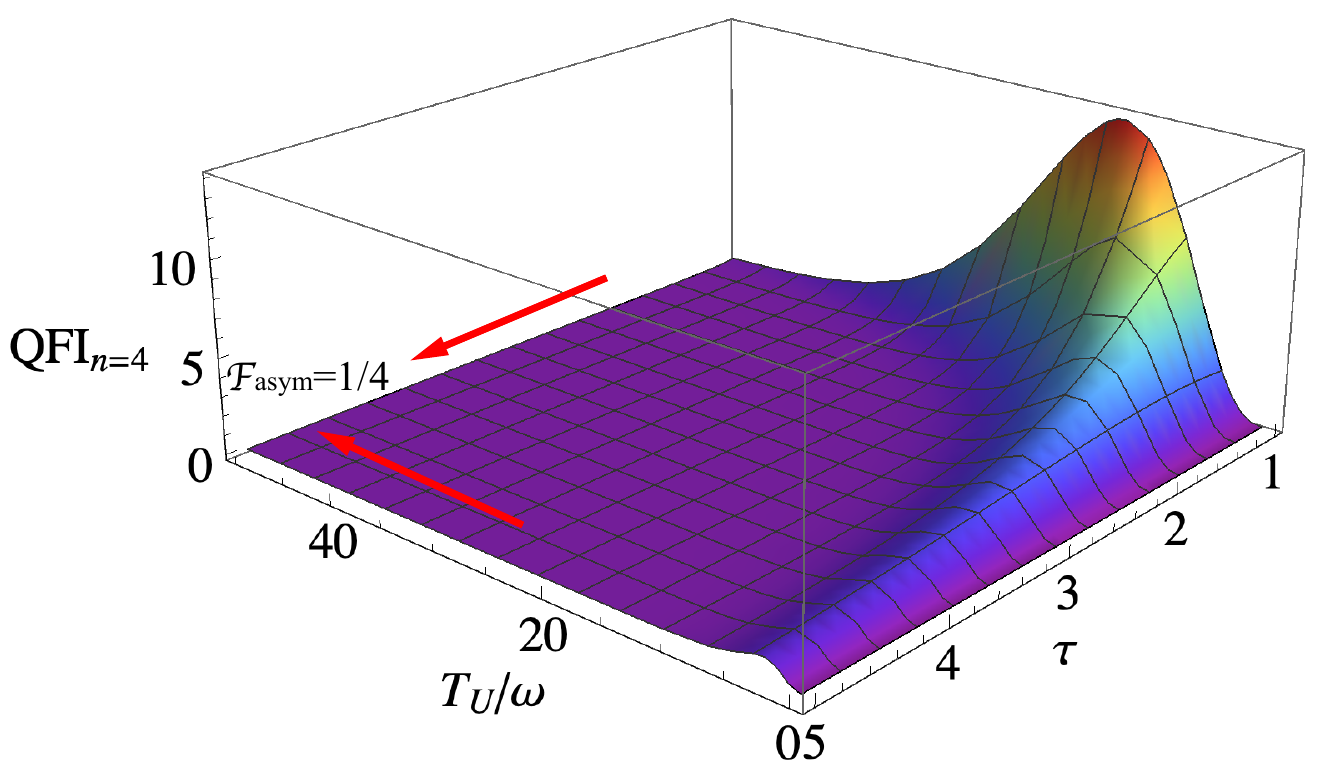}}\\
\subfloat[$\mathcal{F}_{\beta}(n=5)$]{\includegraphics[width=.245\textwidth]{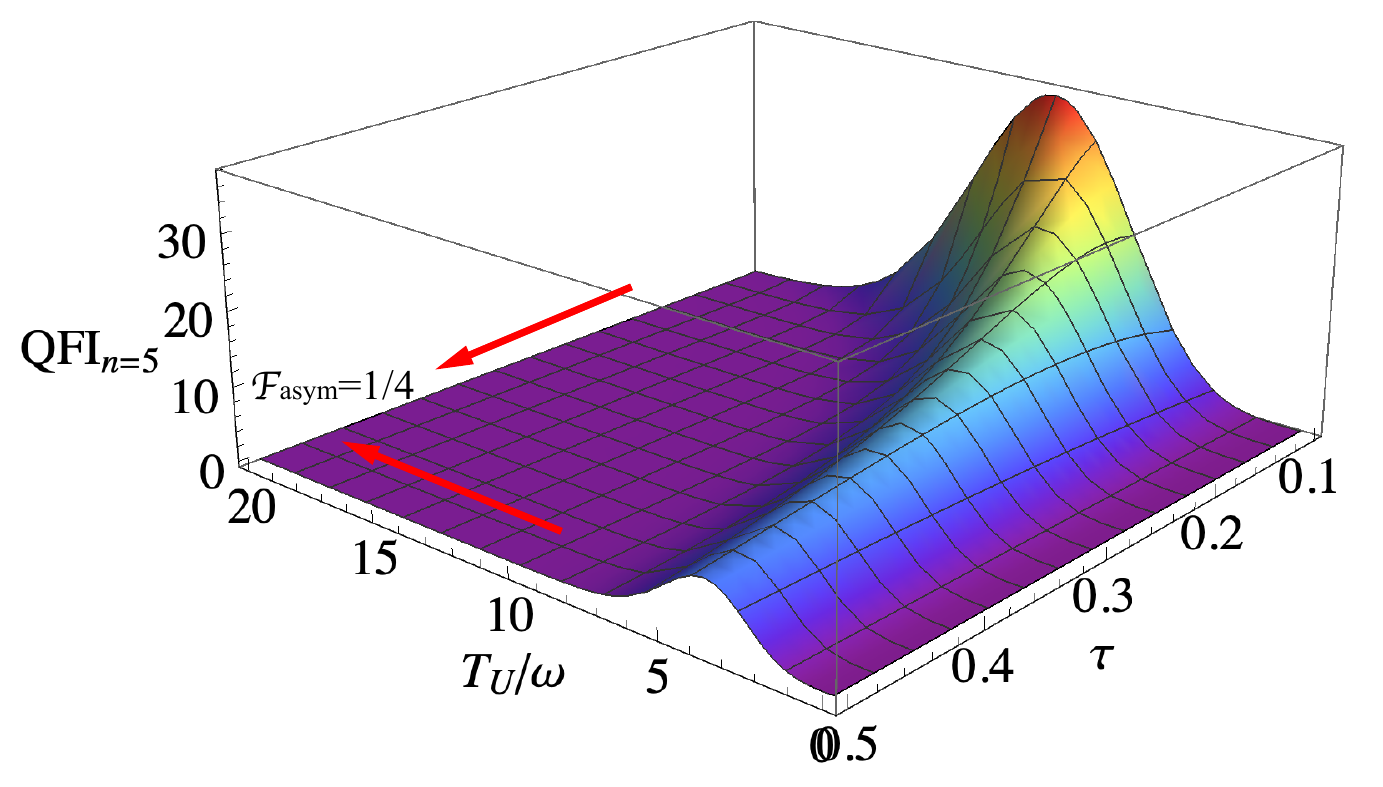}}
\subfloat[$\mathcal{F}_{\beta}(n=6)$]{\includegraphics[width=.24\textwidth]{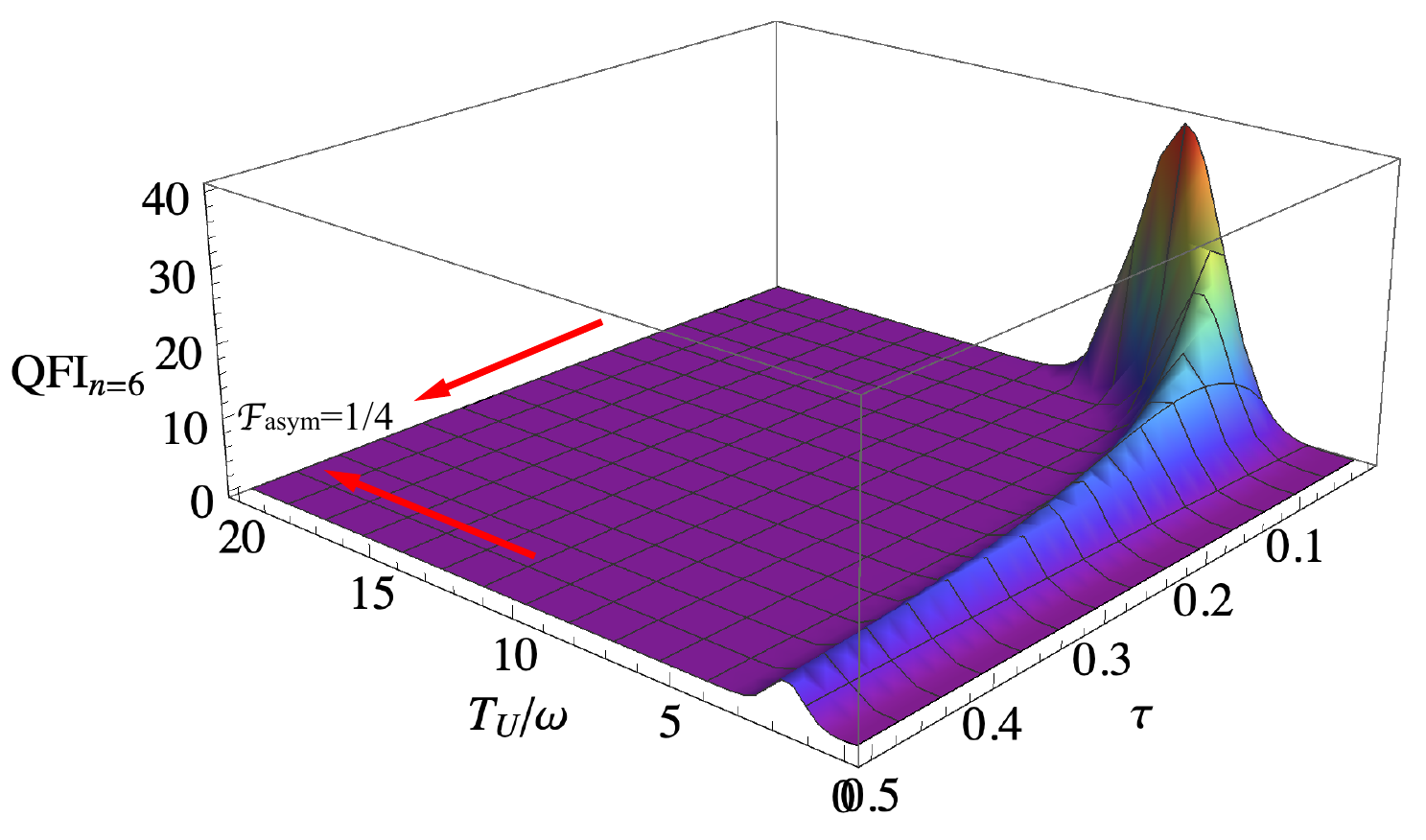}}
\caption{The QFI for the accelerating UDW detector coupled with a massless scalar background in $n=3,4,5,6$-dimensional Minkowski spacetime, as a function of dimensionless time parameter $\tau$ and temperature parameter $T_U/\omega$. For large acceleration and long time evolution, the QFI approaches an asymptotic value $\mathcal{F}_{\text{asym}}=1/4$.}
\label{fig2}
\end{center}
\end{figure}

Without loss of generality, we work in particular cases with $n=3,4,5,6$ and substitutes the Kossakowski coefficients (\ref{eq24}) into (\ref{eq28}). We depict the time-evolving QFI as a function of Unruh temperature in Fig.\ref{fig2}.

The particularity of $n=3$ model immediately steps out via an unique behavior of its QFI. This can be ascribed to the $\beta$-independence of the Kossakowski coefficient, i.e., $\p_\beta\gamma_{+,3}=0$, which simplify (\ref{eq28}) into a monotonic function of $\beta$ and proper time $\tau$. With growing time and increasing Unruh temperature, the QFI increases and monotonically approaches an asymptotic value, e.g., $\mathcal{F}_{\text{asym}}=0.25$. In other word, for $n=3$ case, the optimal precision of estimation can only be achieved for both long enough evolution time and extreme large acceleration.

For the detector moving in $n=4,5,6$ dimension, however, we find that the QFI has a peak value located at early time and finite acceleration, and be significantly larger than their asymptotic value. This suggests that one may achieve an enhanced precision of estimation on Unruh temperature at a relative low acceleration. As we mentioned, such non-monotonicity of the QFI is inherited directly from the polynomial factor on $\beta$ appearing in Kossakowski coefficient (\ref{eq25}), therefore hold for all models with massless scalar background in $n\neq3$ dimension. 

\begin{figure}[hbtp]
\begin{center}
\includegraphics[width=.44\textwidth]{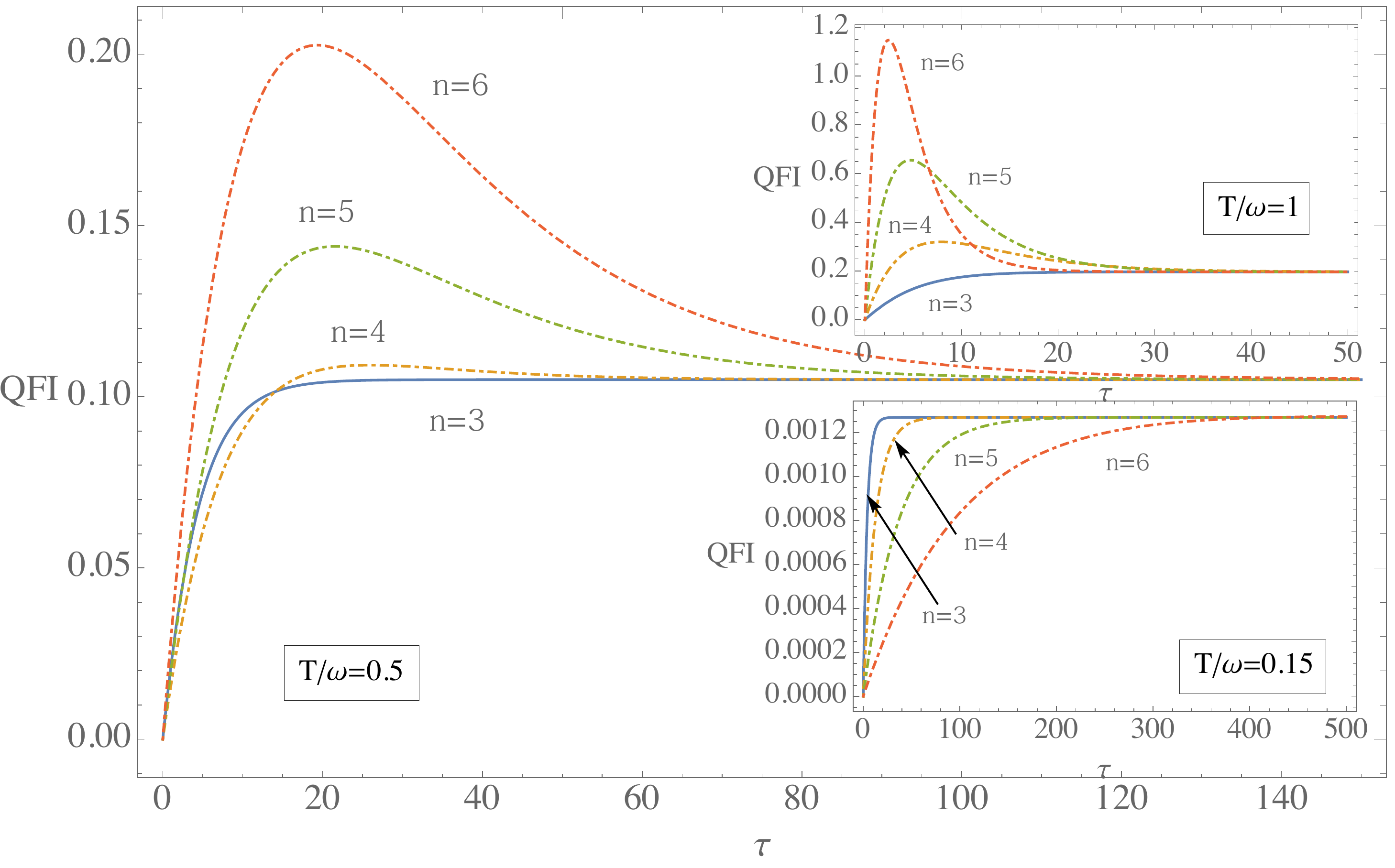}
\caption{The peaks of the QFI for the accelerating detector in $n=3,4,5,6$ dimensional Minkowski spacetime. The time-evolution of QFI for specific dimensionless parameter $T_U/\omega=0.15,0.5,1$ are given respectively. Within higher dimensional spacetime, the QFI has a bigger maximal value which is approached in an earlier time.}
\label{fig3}
\end{center}
\end{figure}

It is interesting to note that the location and magnitude of the QFI peak value can be used to distinguish the different spacetime dimension. We illustrate this in Fig.\ref{fig3} by comparing the refined evolution of QFI in various spacetime dimension. Apart from the striking monotonic behavior in $n=3$, we see that the QFI in higher dimensional spacetime is more sensitive to the Unruh temperature and arrives at its maximum at a relative earlier time, which indicates a more sharp peak during its time evolution. Numerically analysis also shows that the peak value of QFI would increase as the spacetime dimension growing.

Finally, comparing the QFI evolution and detector's response (\ref{eq18}), a crucial difference between them need to be highlighted, where no inversion behavior of QFI as a function of dimensionality was found. This may not be surprising since the QFI evolution essentially distinguishes the way of the detector thermalizing, which is irrelevant with spurious interchange of statistics encoded in the functional form of response function \cite{TAKAGI3}.

\subsection{QFI and mass effect}

For a massive scalar background, the Kossakowski coefficients (\ref{eq22}) is mass-dependent, indicating that the related QFI of an accelerating detector may encode somewhat mass effect of the field. 

The first mass effect we note is that, with a factor $e^{-m\beta/\pi}$ appearing, the coefficients $\gamma_{+,~n}^{\text{massive}}$ are always modulated by a polynomial on $\beta$. Therefore, we conclude that the QFI should now be a non-monotonic function of $\beta$ for arbitrary spacetime dimension, i.e., no such monotonic QFI as shown in Fig.\ref{2}.(a) within massless background can be found once the field acquires mass. 

\begin{figure}[hbtp]
\begin{center}
\includegraphics[width=.45\textwidth]{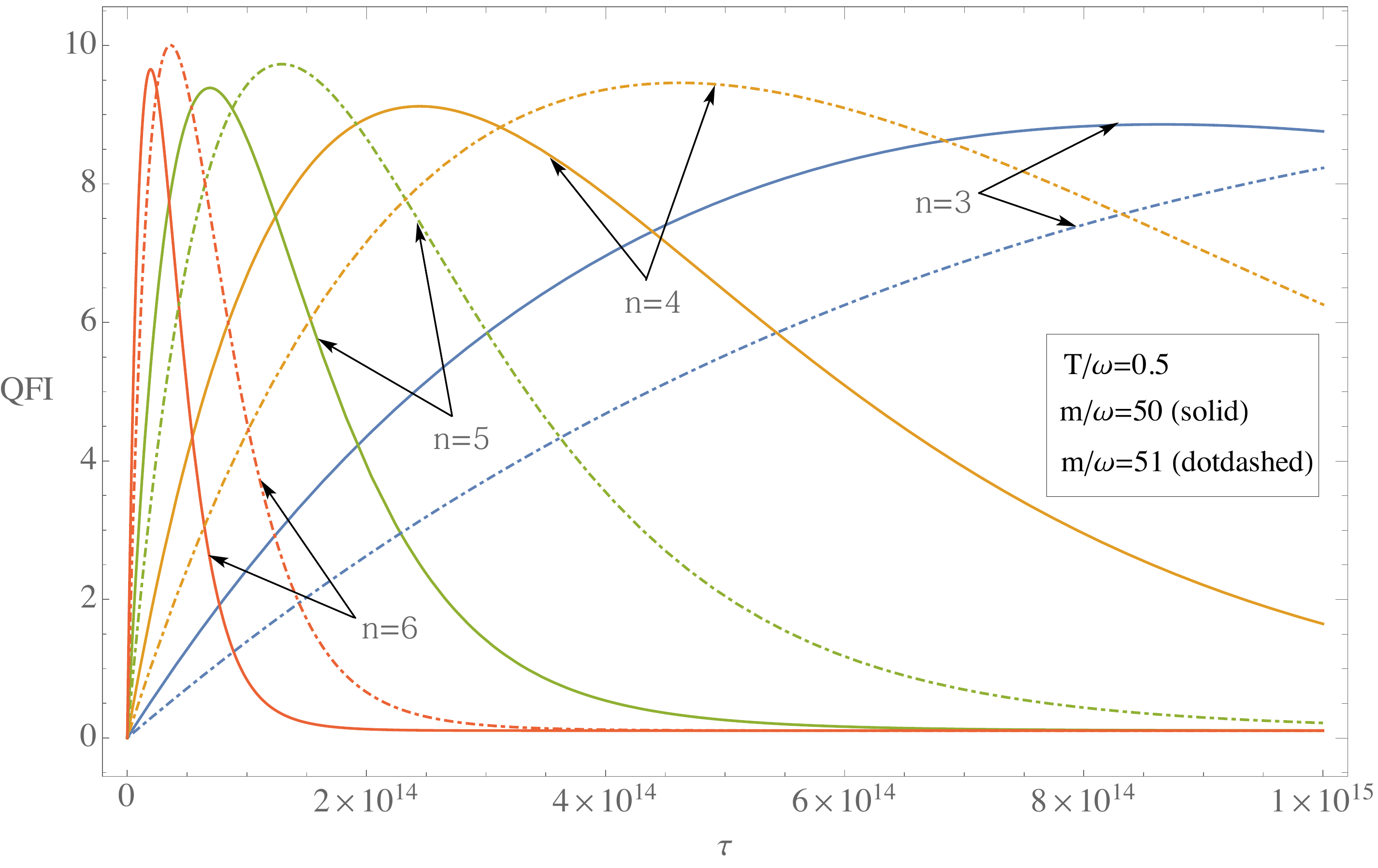}
\caption{The robust QFI for the accelerating detector coupled with a massive background in $n=3,4,5,6$ dimensional spacetime. The QFI has peak value located at early time, which increases for the growing spacetime dimension. For the scalar background with larger mass, the QFI is significantly enhanced.}
\label{fig5}
\end{center}
\end{figure}

To further explore the mass effect encoded in (\ref{eq28}), we depict the time-evolving QFI for $n=3,4,5,6$ massive background with fixed Unruh temperature in Fig.\ref{fig5}. Since analysis should be enrolled in the large field mass limit $m\gg T_U$, we restrict the estimation with low Unruh temperature, e.g., $T_U/\omega=0.5$, $m/\omega=50$. We observe that the non-monotonic QFI can reach a maximal value much larger than the asymptotic value $\mathcal{F}_{\text{asym}}=0.25$, indicating a significant enhancement of the precision of quantum estimation. Similar as in massless background, as the spacetime dimension growing, the QFI has a more sharp peak which has a larger maximum located at an earlier time. 

Nevertheless, it is interesting to note that the mass-dependent QFI can be very robust against the environment decoherence, as its peak sustain for a significantly longer time than it in massless case, whenever the quantum estimation process can be implemented. This is  ascribed to the fact that the response of the UDW detector is now suppressed by an exponential factor of the mass, which makes the detector attain to its equilibrium in a very long time \cite{TAKAGI1}. To demonstrate this, we have compared the QFI for the backgrounds with different mass in Fig.\ref{fig5}, by plotting the case of dimensionless parameter $m/\omega=50,51$ with solid and dot-dashed line respectively. We find that, for a more massive field, the persistence of the related QFI is strengthened while its maximal value increase.

\section{Conclusions}
\label{4}

We have reinvestigated the thermal nature of Unruh effect by using quantum Fisher information as an effective probe. We treat an accelerating UDW detector as an open quantum system coupled to $n$-dimensional scalar background, whose complete dynamics is resolved from a Lindblad master equation of density matrix. The related time-evolving QFI (\ref{eq28}) depends on detector's energy gap, the Unruh temperature $T_U$, as well as the features of the scalar background, such as its mass and dimensionality. We find that the asymptotic QFI whence detector arrives equilibrium is irrelevant to the local response of the detector but only determined by the Unruh temperature $T_U$, thus encapsulates the global side of Unruh thermality guaranteed by the KMS condition. On the other hand, we show that the local side of Unruh effect, i.e., the different ways to approach the same thermal equilibrium, is encoded in the time-evolution of the related QFI. Within massless scalar background, the QFI has unique monotonicity in $n=3$ dimensional spacetime, and becomes non-monotonous for $n\neq3$ models, where a local peak value at finite acceleration can exist. In a metrological view, this means that an enhanced precision of estimation on Unruh temperature can be achieve at a relative low acceleration. Finally, we emphasized that once the field acquiring mass, the related QFI is robust against the Unruh decoherence, in the sense that its local peak can sustain for a very long time. For the background field with larger mass, the persistence of the QFI can even be significantly strengthened with larger maximal value. Such robustness of QFI can surely facilitate any practical quantum estimation task.

Our results agree with other previous studies \cite{TAKAGI5,TAKAGI6,TAKAGI7}, claiming that the thermal nature of Unruh effect is correct and has nothing to do with the particularities of detector's local response. With utilizing of QFI, we can further refine the physical information extracted from the local response as the maximal accuracy of possible quantum estimation process \cite{FI5}. It would be interesting to extend our study to other background field, like the fermionic fields considered in \cite{TAKAGI3} or those related to intermediated statistics \cite{TAKAGI9}, where the analysis presented in this Letter can apply. 

\section*{Acknowledgement}
This work is supported by the National Natural Science Foundation of China (No.12075178). J. F. would like to thank Yao-Zhong Zhang for enlightening discussions and kind hospitality at the University of Queensland where part of this work started.

\end{document}